# Stress-induced modification of gyration dynamics in stacked double-vortex structures studied by micromagnetic simulations

Vadym Iurchuk[1], Attila Kákay[1] and Alina M. Deac[2]

[1] Institute of Ion Beam Physics and Materials Research, Helmholtz-Zentrum Dresden-Rossendorf, 01328 Dresden, Germany

[2] Dresden High Magnetic Field Laboratory, Helmholtz-Zentrum Dresden-Rossendorf, 01328 Dresden, Germany

**Abstract**

In this paper, using micromagnetic simulations, we investigate the stress-induced frequency tunability of double-vortex nano-oscillators comprising magnetostrictive and non-magnetostrictive ferromagnetic layers separated vertically by a non-magnetic spacer. We show that the the relative orientations of the vortex core polarities $p_1$ and $p_2$ have a strong impact on the eigen-frequencies of the dynamic modes. When the two vortices with antiparallel polarities have different eigen-frequencies and the magnetostatic coupling between them is sufficiently strong, the stress-induced magnetoelastic anisotropy can lead to the single-frequency gyration mode of the two vortex cores. Additionally, for the case of parallel polarities, we demonstrate that for sufficiently strong magnetostatic coupling, the magnetoelastic anisotropy leads to the coupled vortex gyration in the stochastic regime and to the lateral separation of the vortex core trajectories. These findings offer a fine control over gyration frequencies and trajectories in vortex-based oscillators via adjustable elastic stress, which can be easily generated and tuned electrically, mechanically or optically.

Corresponding author's e-mail: v.iurchuk@hzdr.de

## 1. Introduction

The ground state of nanoscale circular magnetic disks of certain geometric aspect ratios is a spontaneously forming stable vortex configuration with circulating in-plane magnetization and a vortex core pointing out-of-plane. Resonantly exciting the vortex core via either an rf magnetic field or an rf spin-polarized current yields a gyrotropic motion around its equilibrium position, characterized by a specific eigen-frequency. The gyrotropic frequency depends on the material parameters and the disk geometry [1]. Such oscillations, which can be read out via periodic magnetoresistance changes, generate rf signals with output powers in the nW range and linewidth below 1 MHz, centered at the frequencies of few hundred MHz, thus yielding high quality factors (~1000) [2]. The quality factor can be further improved by one order of magnitude when the coupled vortex mode is excited in stacked double-vortex structure [3,4]. Subsequently, a complex dynamics of coupled vortex modes was studied in details in [5] revealing fine splitting of the corresponding resonant frequencies. Recent demonstration of the phase noise reduction in vortex-based spin-torque oscillators locked to an external rf current [6,7] provides the potential for system-level integration of vortex-based devices for such applications as spectrum analysis [8,9], microwave receivers/transmitters [10] and spintronics-based neuromorphic computing [11,12]. While all these features make vortex-based nano-oscillators interesting as nanoscale rf sources, the major drawback remains their low frequency tunability when operated in linear regime.

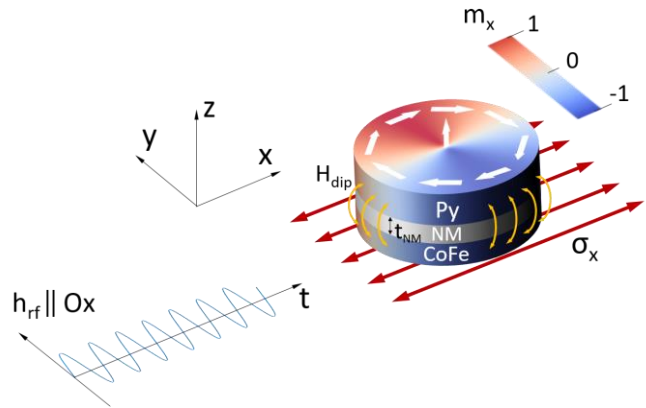

Fig. 1. Schematics of the simulated experiment. The in-plane sinusoidal rf magnetic field is applied along the x-axis to the CoFe/NM/Py double-vortex structure to excite the gyration dynamics of the CoFe and Py vortex cores coupled magnetostatically via dipolar magnetic field $H_{dip}$. The structure is biased by a uniaxial tensile stress $\sigma_x$ along the x-axis.

Recently, the magnetoelastic effect has been proposed to enhance the tunability of the eigen-frequencies of individual vortices. It was shown both theoretically [13] and experimentally [14] that introducing an additional magnetoelastic anisotropy term to the vortex core dynamics leads to the softening of the restoring force constants and thus to the gyrotropic frequency decrease. This enables the stress-induced controlling of the vortex eigen-frequency e.g. via bending of the flexible membrane-like substrate [14], an application of an electric field to the piezoelectric substrate [15,16] or even using optically generated deformation of photostrictive materials [17,18].

Here, we present a micromagnetic study of the interplay between the strain-induced magnetic anisotropy and magnetostatic interaction in vortex pairs separated vertically by a non-magnetic spacer. Specifically, we consider a double-disk structure comprising magnetostrictive (CoFe) and non-magnetostrictive (Py) ferromagnetic layers separated vertically by a non-magnetic spacer. We show that, when the two vortices with anti-paralel polarities have different eigen-frequencies and the magnetostatic coupling between them is sufficiently strong, the stress-induced magnetoelastic anisotropy can lead to the single-frequency gyration mode of the two vortex cores. On the contrary, for the case of parallel vortex core polarities, the stress-induced magnetoelastic anisotropy affects considerably not only the resonant frequencies but also the vortex core gyration trajectories leading to the stochastic dynamics of the magnetostaically coupled vortex pair. These findings offer a frequency tunability of double-vortex-based oscillators via elastic stress, which can be generated and controlled electrically, mechanically, or optically.

## 2. Simulations details

In this paper, we report on the simulations of the vortex dynamics in double-vortex structure in the presence of magnetostatic coupling and stress-induced magnetoelastic anisotropy. We use the GPU-accelerated MuMax3 software package [19] for the simulations of the magnetization dynamics in nanosized magnetic disks under the presence of the stress-induced magnetoelastic anisotropy. We consider a tri-layer system (see Fig 1) comprising disk-shaped magnetostrictive (MS) and non-magnetostrictive (nMS) ferromagnetic layers separated by a non-magnetic (NM) spacer with the thickness $t_{NM}$ providing magnetostatic coupling between the magnetic layers. We study the evolution of the frequency and trajectory of the vortex core gyrotropic mode as a function of the in-plane uniform uniaxial stress $\sigma$ imposed onto the tri-layer. Due to the inverse magnetostrictive effect, the stress $\sigma$ introduces an additional anisotropy energy in the form of $K_\sigma = \frac{3}{2}\lambda_s\sigma$, where $\lambda_s$ is the magnetostriction constant of the magnetic material [20]. When $\lambda_s$ is positive (negative), a tensile stress imposes a magnetoelastic anisotropy along (perpendicular to) the direction of the applied stress.

In our simulations, we choose CoFe as MS material providing significant magnetostriction constant $\lambda_s^{CoFe} = 65\cdot 10^{-6}$ [21,22], and Permalloy (namely $Ni_{80}Fe_{20}$ or Py) as nMS material known to have a negligible magnetostriction $\lambda_s^{Py} \approx 0$ [23]. We use the following material parameters for the magnetization dynamics simulations: saturation magnetizations $M_s^{CoFe} = 1700$ kA/m for CoFe and $M_s^{Py} = 800$ kA/m for Py, exchange constants $A_{ex}^{CoFe} = 21$ pJ/m³ for CoFe and $A_{ex}^{Py} = 13$ pJ/m³ for Py, damping constant $\alpha = 0.008$ for both layers. For all simulated data presented here, we consider the MS/NM/nMS tri-layer of 300 nm diameter discretized into rectangular prisms with the size of 2.34 nm in the x and y directions and 10 nm in the z-direction. The magnetic vortex cores are excited by an in-plane sinusoidal rf magnetic field $\bm{h}_{rf} = \bm{h}\sin(2\pi f t)$, where $f$ is the excitation frequency, $\bm{h} = (h_x; 0; 0)$ and the amplitude $h_x = 1$ mT. This relatively small magnetic field amplitude is chosen in order to drive the vortex core dynamics within the linear gyration regime. The dynamic magnetization distributions in both magnetic disks are simulated during the time $T$ of 60 periods ($T = \frac{60}{f}$), and the final magnetization state of each disk is captured for each value of the excitation frequency $f$ in the chosen range. Thus, we obtain an FMR-like absorption spectrum with the resonance peak corresponding to the eigen-frequency of the vortex core gyration. We show that in such double-vortex structure, the resonance frequencies of the individual and coupled modes, which in general are defined by the vortex polarities and the spacer thickness (i.e. magnetostatic coupling strength), can be modified by introducing the magnetoelastic energy $K_\sigma$.

## 3. Results and discussion

### 3.1. Vortex core dynamics in CoFe/NM/Py double-vortex structures: spacer thickness dependence

An individual vortex is featured by the in-plane magnetization circulation c and the core polarity p, defining four distinct vortex configurations: $(c;p) = (1;1)$, $(c;p) = (1;-1)$, $(c;p) = (-1;1)$ and $(c;p) = (-1;-1)$. The eigen-frequencies of the vortex core gyration are mainly defined by the magnetic parameters of the material (i.e. saturation magnetization, exchange constant, anisotropies, etc) and the geometrical thickness-to-diameter ratio of the circular disk [1] and do not depend on the $(c;p)$ configuration. Similarly, for the case of double-vortex structure, four non-degenerate configurations of $c$ and $p$ can be distinguished: $(c_1c_2;p_1p_2) = (1;1)$, $(c_1c_2;p_1p_2) = (1;-1)$, $(c_1c_2;p_1p_2) = (-1;1)$ and $(c_1c_2;p_1p_2) = (-1;-1)$. However, the dynamics of the double-vortex structure is more complex and depends not only on the material parameters and geometry, but also on the mutual orientation of the vortex polarities and circulations as well as on the efficiency of the coupling between the vortices. In this section, we discuss the resonant dynamics of the CoFe/NM/Py double-vortex structure, excited by the in-plane rf magnetic field $h_{rf} = h_x sin(2\pi ft)$ as a function of the NM spacer thickness and for different $(c_1c_2;p_1p_2)$ configurations.

#### 3.1.1. Antiparallel polarities

Fig. 2(a) shows the simulated FMR spectra of the CoFe(10)NM($t_{NM}$)Py(20) cylindrical tri-layer (thicknesses are in nm, diameter $d$ = 300 nm) as a function of the excitation frequency $f$ for different thicknesses $t_{NM}$ of the NM layer ranging from 10 to 160 nm. Here we consider the case of antiparallel polarities and circulations, i.e $(c_1c_2;p_1p_2) = (-1;-1)$. All spectra feature double-resonance behavior with low- (high-) frequency peak $f_{lo}$ ($f_{hi}$) corresponding to the Py (CoFe) layer resonance (Fig. 2(c)), which is confirmed by the simulations of the vortex core trajectories of the corresponding layer (Fig. 2(d,e)). At $f_{lo}$ ($f_{hi}$) the Py (CoFe) resonant trajectory is quasi-circular and corresponds to the steady-state vortex core gyration. On the contrary, the CoFe (Py) trajectory exhibits periodic low-amplitude off-resonance deflection of the vortex core from the equilibrium position.

Interestingly, a slight deformation of the both trajectories is present for the two resonance frequencies, which is attributed to the non-zero magnetostatic coupling between the two layers.

For $t_{NM} \geq 200$ nm, the resonances occur at $f_{lo} = f_{Py} = 0.56$ GHz and $f_{hi} = f_{CoFe} = 0.62$ GHz corresponding to the resonances of the uncoupled CoFe(10) and Py(20) disks, respectively. Decreasing the spacer thickness leads to a redshift of the eigen-frequencies of both layers as well as to a decreased frequency separation $\Delta f \equiv f_{hi} - f_{lo}$ due to the increased magnetostatic coupling between the magnetic layers, pulling the frequencies closer to each other.

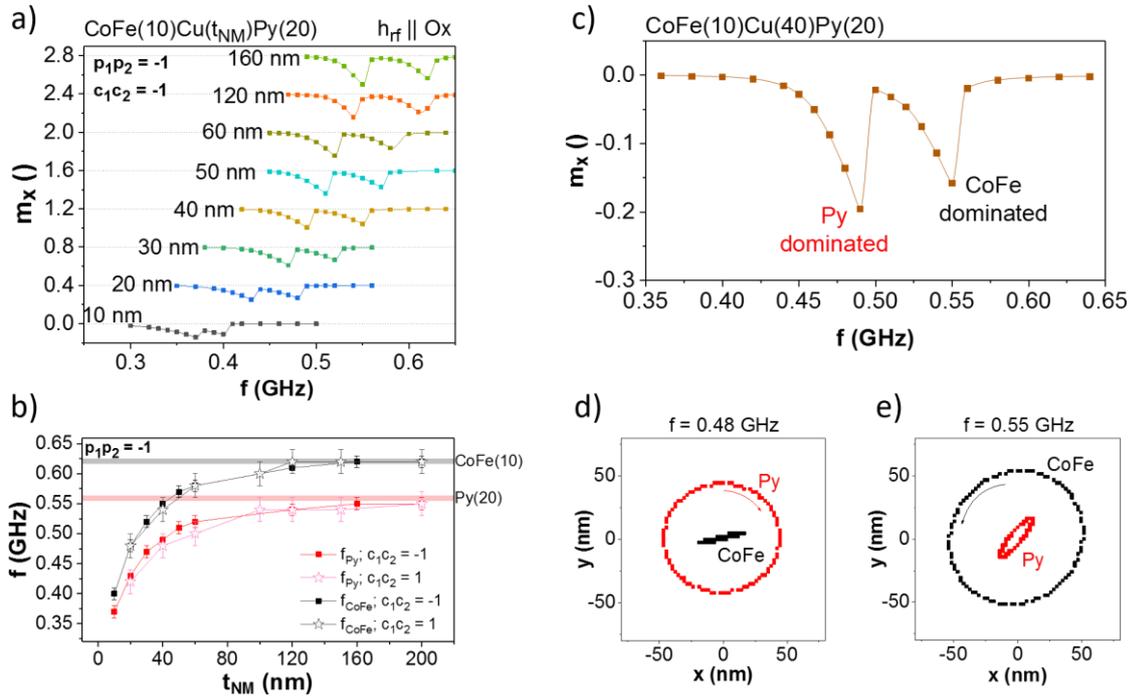

Fig. 2. (a) Simulated FMR spectra of the CoFe(10)NM($t_{NM}$)Py(20) double vortex structure as a function of spacer thickness $t_{NM}$ for the $(c_1 c_2; p_1 p_2) = (-1; -1)$ case. The plots are vertically offset for better visibility. (b) Resonant frequencies as a function of $t_{NM}$ for the two combinations of in-plane circulations $c_1$ and $c_2$. Open stars (solid squares) are used to denote parallel (anti-parallel) circulation directions. Horizontal bars correspond to the eigen-frequencies of the isolated CoFe(10) and Py(20) disks. (c) FMR spectra of the CoFe(10)NM(40)Py(20) tri-layer featuring Py-dominated and CoFe-dominated resonances. Lines in (a-c) are guides to the eye. (d,e) Simulated trajectories of the CoFe (black) and Py (red) vortex cores recorded at the resonant excitation frequencies of the plot (c). The trajectories were recorded after 100 gyration periods to ensure that the steady-state dynamics is set.

The resonant dynamics of the double-vortex structure is mostly defined by the mutual orientation of the core polarities. As seen from Fig. 2(b), changing the circulation direction in one of the magnetic layers introduces only minor changes to the resonant frequencies of the magnetostatically coupled disks. The direction of the gyrotropic motion, according to the Thiele's equation [1,24], is defined by the vortex core polarity and, for our geometry, is counter-clockwise for the CoFe layer resonance ($p_1 = 1$) and clockwise for the Py layer resonance ($p_2 = -1$).

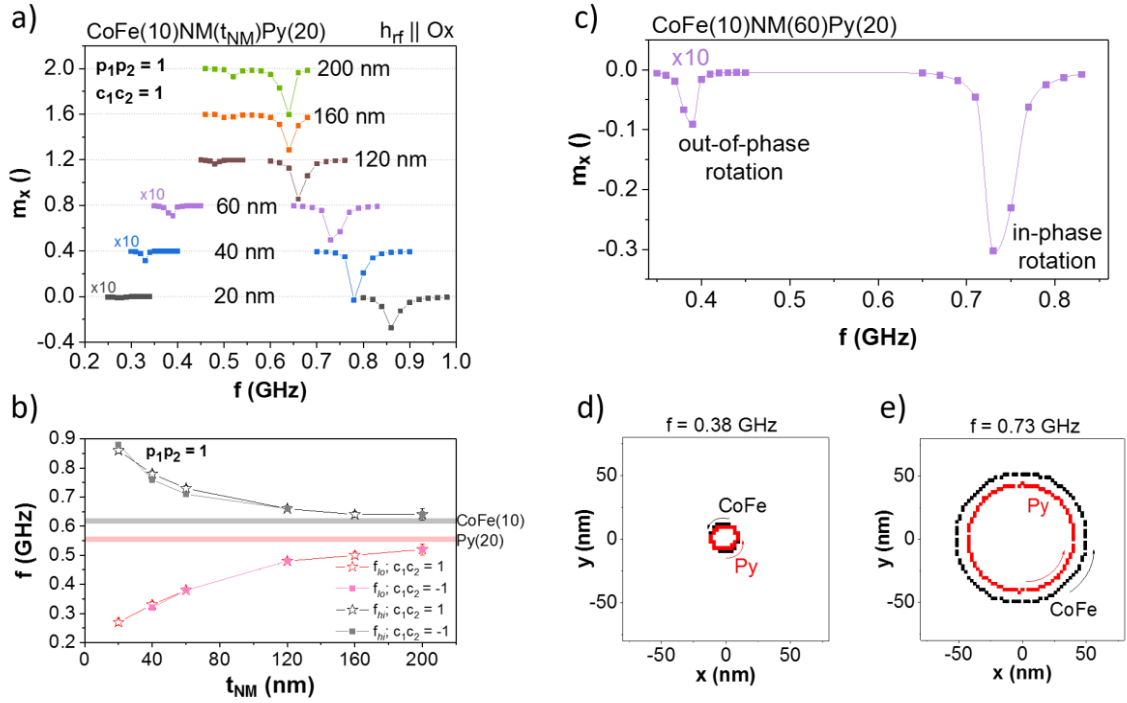

Fig. 3. (a) Simulated FMR spectra of the CoFe(10)NM($t_{NM}$)Py(20) double-vortex structure as a function of spacer thickness $t_{NM}$ for the $(c_1 c_2; p_1 p_2) = (1; 1)$ case. The plots are vertically offset for better visibility. The amplitudes of the low-frequency branch for $t_{NM} = 20; 40$ and $60$ nm are multiplied by a factor of 10. (b) Resonant frequencies as a function of $t_{NM}$ for different combinations of in-plane circulations $c_1$ and $c_2$ denoted by open stars (solid squares) for parallel (anti-parallel) circulation directions. Horizontal bars correspond to the eigen-frequencies of the isolated CoFe(10) and Py(20) disks. (c) FMR spectra of the CoFe(10)NM(60)Py(20) tri-layer featuring low-frequency out-of-phase gyration and high-frequency in-phase gyration resonances. Lines in (a-c) are guides to the eye. (d,e) Simulated trajectories of the CoFe (black) and Py (red) vortex cores recorded at the resonant excitation frequencies of the plot (c).

### 3.1.2. Parallel polarities

For the case of parallel polarities ($p_1 p_2 = 1$), the evolution of the resonant dynamics of the double vortex structure as a function of t$_{NM}$ is qualitatively dissimilar from the $p_1 p_2 = 1$ case. Fig. 3(a) shows the simulated dynamical spectra of the CoFe(10)NM($t_{NM}$)Py(20) tri-layer ($d = 300$ nm) for different $t_{NM}$ for the $(c_1 c_2; p_1 p_2) = (1; 1)$ case. Similar to the $p_1 p_2 = -1$ case, the dynamics of the double-vortex structure corresponds to the resonant gyration of the uncoupled isolated CoFe(10) and Py(20) vortices when the magnetostatic coupling is negligible, i.e. the NM spacer is thick ($t_{NM} \geq$ 200 nm). Contrary to the $p_1 p_2 = -1$ case, when the spacer thickness is decreased, i.e for $t_{NM} \lesssim 160$ nm, the high-frequency resonance $f_{hi}$ is blue-shifted towards higher frequencies, whereas low-frequency resonance $f_{lo}$ is red-shifted towards lower frequencies thus increasing the frequency separation $\Delta f$. The frequencies of the excited dynamic modes are defined by the vortex core polarities with only marginal contribution from the in-plane circulations $c_1$ and $c_2$ of the vortices (see Fig. 3(b)).

In the case of parallel polarities, the increased magnetostatic interaction leads to the "repulsion" of the frequencies of the excited modes and to the decreased amplitude of the low-frequency mode. For example, at $t_{NM} = 60$ nm, the low-frequency mode amplitude is more than one order of magnitude smaller as compared to the amplitude of the high-frequency mode (see Fig. 3(c)). By examining the resonant trajectories of the corresponding modes, we attribute the low-frequency mode to the so-called small amplitude "out-of-phase" gyration of the CoFe and Py vortex cores, which both gyrate counter-clockwise with the constant lateral separation (see Fig. 3(d)), i.e. their azimuthal components are $\pi$-shifted. The high-frequency mode corresponds to the large amplitude "in-phase" resonant gyration where both vortex cores gyrate with the same azimuthal components (see Fig. 3(d)). The observed dynamics is in full agreement with the analytical and micromagnetic description provided in [25] for the case of magnetostatically coupled symmetric of the double-vortex structure with parallel core polarities.

Hence, the dynamics of two stacked vortices coupled magnetostatically features a complex double resonance behavior with the possibility to manipulate the excited mode type and frequency by adjusting the geometry of the structure (i.e. disk aspect thickness-to-diameter aspect ratio and the spacer thickness) and by properly initiating the mutual orientation of the vortex core polarities.

### 3.2. Stress-tunable gyration dynamics in CoFe/NM/Py double-vortex structures

In this section, we present the stress-dependent tuning of the resonant frequencies and trajectories of the magnetostatically coupled MS/NM/nMS double-vortex structure. We introduce the magnetoelastic energy $K_\sigma = \frac{3}{2}\lambda_s\sigma$ to the dynamic simulations of the gyration of the MS vortex core (here CoFe). We show that, for $p_1p_2 = -1$ configuration of the vortex polarities, the stress-induced magnetoelastic anisotropy can switch the system from the double-resonance to the single-resonance regime. Moreover, for the case of parallel polarities ($p_1p_2 = 1$) and strong magnetostatic coupling, we demonstrate the stress-induced lateral shift of the vortex core gyration trajectories.

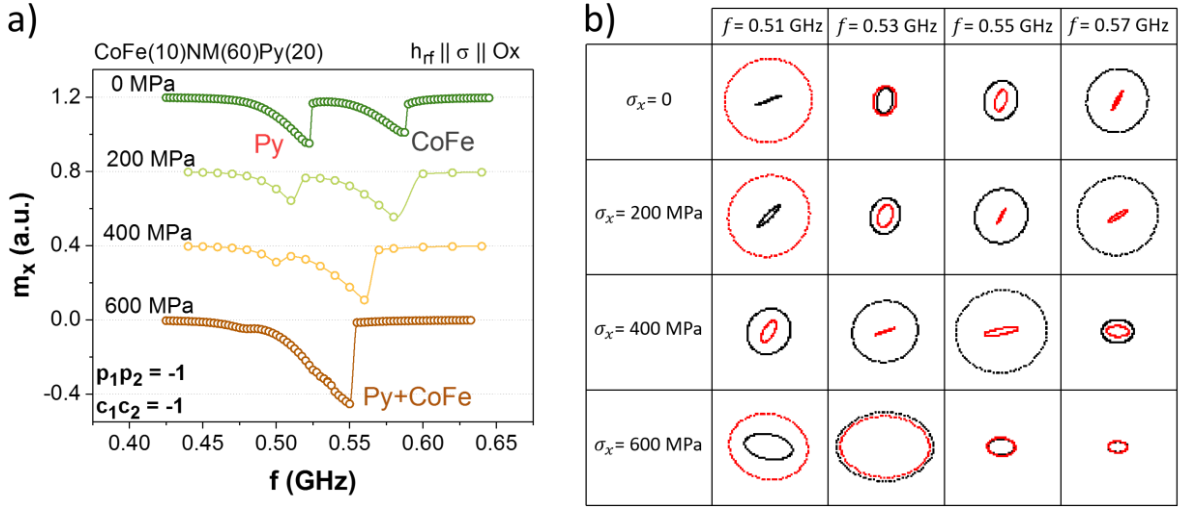

Fig. 4. (a) Simulated spectra of the 300 nm diameter CoFe(10)NM(60)Py(20) double-vortex structure for the $(c_1c_2; p_1p_2) = (-1; -1)$ case taken at different uniaxial tensile stress $\sigma_x$. The plots are vertically offset for better visibility. (b) Trajectories of the CoFe (black) and Py (red) vortex cores recorded for excitation frequencies $f = 0.51; 0.53; 0.55$ and $0.57$ GHz without applied stress ($\sigma_x = 0$) and for $\sigma_x = 200; 400$ and $600$ MPa. The table cell size corresponds to the square area of $80\times80$ nm$^2$ centered in the middle of the disk.

#### 3.2.1. Antiparallel polarities

Fig. 4(a) shows the simulated spectra of the CoFe(10)NM(60)Py(20) double vortex structure for the $(c_1c_2; p_1p_2) = (-1; -1)$ case and for different values of the uniaxial tensile stress $\sigma_x$ applied to the tri-layer along the x-axis. The stress introduces an additional magnetoelastic energy $K_\sigma = \frac{3}{2}\lambda_s^{CoFe}\sigma_x$ along the x-axis of the CoFe layer via inverse magnetostrictive effect. Since Py has negligible magnetostriction constant, no stress-induced anisotropy is exerted on the Py layer. As the $\sigma_x$ magnitude increases from 0 to 400 MPa, we observe a clear decrease of the resonance frequencies of both CoFe and Py vortex cores. The decrease of the CoFe resonance frequency at moderate stress values is attributed to the softening of the restoring force spring constants [13] due to the stress-induced additional magnetoelastic anisotropy $K_\sigma$. The Py resonance frequency also follows the decreasing trend, as it is magnetostatically coupled to the CoFe layer via the spacer. Notably, the amplitude of the CoFe (Py) vortex resonance increases (decreases) with increased stress indicating a transition from double-resonance towards CoFe-dominated dynamics. This transition is also confirmed by visualizing the trajectories of the CoFe and Py vortex cores for different values of the applied stress and for different values of the excitation magnetic field frequency (Fig. 4(b)). For $\sigma_x = 0$ and 200 MPa (corresponding to $K_\sigma = 19.5$ kJ/m3), the excitations at 0.51 GHz and 0.57 GHz correspond to the Py and CoFe vortex core resonances respectively separating the frequency ranges of the Py-dominated (lower frequency) and the CoFe-dominated (higher frequency) gyration. For $\sigma_x = 400$ MPa (corresponding to $K_\sigma = 39$ kJ/m3), an onset of the transition towards the CoFe-dominated dynamics is observed, where CoFe is at resonance around $f = 0.56$ GHz (see green graph in Fig. 4(a)), while the gyration of the Py vortex core is suppressed as evidenced by the gyration trajectories of Fig. 4(b). Notably, when the stress is further increased up to 600 MPa (corresponding to $K_\sigma = 58.5$ kJ/m3), the dynamics of the double-vortex features single-resonance behavior with increased amplitude of the resonant peak (blue graph in Fig. 4(a)). This resonance is attributed to the synchronized gyration of both CoFe and Py vortex cores, which is confirmed by examining the gyration trajectories (Fig. 4(b)).

The observed transition from the double-resonance to the single-resonance dynamics can be qualitatively understood as follows. For the magnetostatically coupled MS/NM/nMS double-vortex structure, the frequency splitting is defined by the resonant eigen-frequencies of the isolated vortices and the strength of the interlayer magnetostatic coupling (as shown in section 33.1.1). Stress-induced magnetoelastic anisotropy leads to the redshifting of the CoFe vortex eigen-frequency. When the magnetostatic coupling between the layers is negligible (i.e. the spacer is thick), both vortices gyrate independently and the redshifting of the CoFe eigen-frequency do not imprints any modification of the Py vortex dynamics (see the simulated dynamics for $t_{NM} = 160$ nm in the Suppl. data). For decreased spacer thickness, the increased magnetostatic interaction becomes sufficient to overcome the separation between the CoFe and Py eigen-frequencies and to bring the system to the large-amplitude gyration of both vortex cores in the synchronized regime.

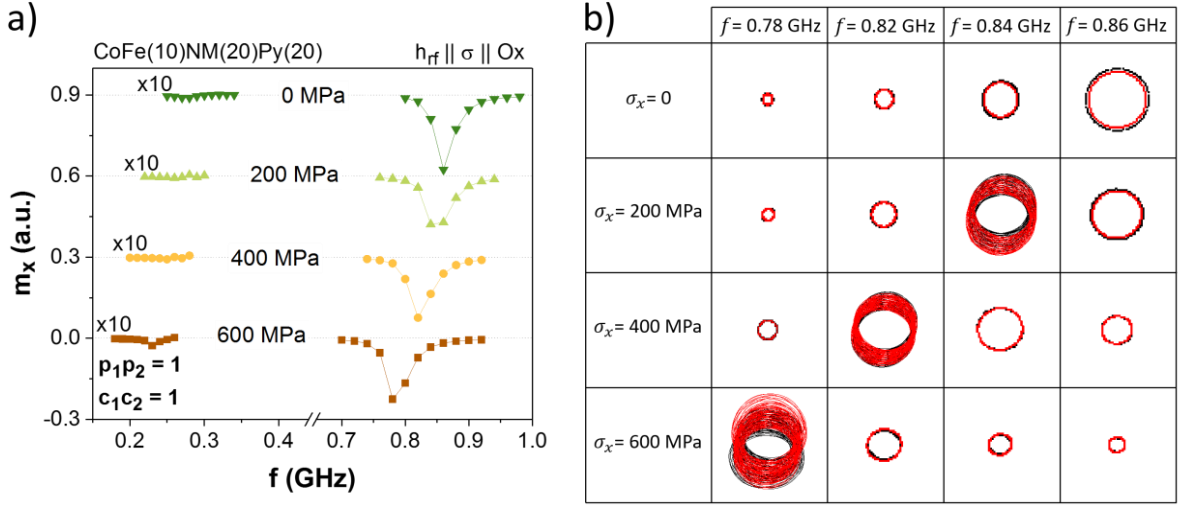

Fig. 5. (a) Simulated spectra of the 300 nm diameter CoFe(10)NM(20)Py(20) double-vortex structure for the $(c_1 c_2; p_1 p_2) = (1; 1)$ case taken at different uniaxial tensile stress $\sigma_x$. The plots are vertically offset for better visibility. The amplitudes of the low-frequency branch are multiplied by a factor of 10. (b) Trajectories of the CoFe (black) and Py (red) vortex cores recorded for excitation frequencies $f = 0.78$; 0.82; 0.84 and 0.86 GHz without applied stress ($\sigma_x = 0$) and for $\sigma_x = 200$; 400 and 600 MPa. The table cell size corresponds to the square area of 80×80 nm² centered in the middle of the disk.

### 3.2.2. Parallel polarities

For the $p_1 p_2 = 1$ case (i.e. for parallel polarities), the dynamics of magnetostatically coupled vortices under stress is qualitatively different from the $p_1 p_2 = -1$. The dynamics features high-frequency large-amplitude in-phase mode and low-frequency small-amplitude out-of-phase mode, as described in details in section III3.1.2. Fig. S2(a) in the Suppl. data shows the simulated spectra of the CoFe(10)NM(60)Py(20) double vortex structure for the $(c_1 c_2; p_1 p_2) = (1; 1)$ case and for different values of the uniaxial tensile stress $\sigma_x$ applied along the x-axis. The spacer thickness of $t_{NM} = 60$ nm corresponds to the moderate magnetostatic coupling (Fig. 3(b)). When the stress increases, a clear redshift of the in-phase gyration mode is observed. This redshift of the coupled CoFe/NM/Py system is driven by the CoFe vortex eigen-frequency decrease due to the softening of the restoring force spring constants in presence of the stress-induced magnetoelastic anisotropy. The low-frequency out-of-phase mode also exhibits redshifting behavior following the frequency "repulsion" behavior for the magnetostatically coupled vortices with parallel polarities. The resonant gyration trajectories of the high-frequency are concentric and have equal azimuthal components over one period (e.g. both vortices gyrate "in-phase") and their circumference and shape depend only marginally on the applied stress (see Fig. S2(b) in the Suppl. data).

However, when the spacer thickness is reduced to $t_{NM} = 20$ nm, an unexpected stress-induced behavior of the resonant trajectories is observed as a result of strong magnetostatic coupling between the layers. While the resonant frequencies follow the redshifting trend (see Fig. 5(a)) (similar to the $t_{NM} = 60$ nm case), the resonant trajectories are significantly modified as the stress increases from 0 to 600 MPa (Fig. 5(b)). Notably, starting from $\sigma_x = 200$ MPa, the trajectories of CoFe and Py vortex cores are no longer concentric but exhibit a lateral separation while conserving the "in-phase" gyration character with equal azimuthal components. Moreover, in the presence of the magnetoelastic anisotropy, both CoFe and Py vortex core trajectories of the in-phase mode deviate from the steady state circular orbits (see Fig. 5(b) for $\sigma_x = 0$) and exhibit low-frequency beatings around the equilibrium point (see Fig. 5(b) for $\sigma_x = 200; 400$ and $600$ MPa). Fig. 6(a) shows the time-varying y-coordinate of the CoFe and Py vortex cores recorded for 100 ns (corresponding to 84 gyration periods at $f = 0.84$ GHz) at $\sigma_x = 200$ MPa. The time traces reveal the periodic quasi-sinusoidal modulation of the circular vortex core trajectory with the beating frequency of ~40 MHz extracted by FFT transform of the time-varying y-coordinate (see Fig. 6(c)). The beatings of CoFe and Py vortex cores are in "anti-phase", i.e. when at certain moment in time the y-coordinate of the CoFe vortex core reaches the maximum (minimum), the corresponding y-coordinate of the Py core is at its minimum (maximum). Thus, the maximum trajectories separation increases from ~32 nm for $\sigma_x = 200$ MPa to ~43.5 nm for $\sigma_x = 600$ MPa. Notably, when the stress is increased up to 600 MPa, the beatings become stochastic without any visible periodic pattern (see Fig. 6(b)). It is obvious that the frequency of these beatings is almost one order of magnitude lower comparing to that of the low-frequency out-of-phase gyration mode $f_{lo}$, therefore one can conclude that these "beatings" essentially differ from the low-frequency variation of the gyration trajectories described in [25] as a result of strong dynamic interlayer magnetostatic interaction.

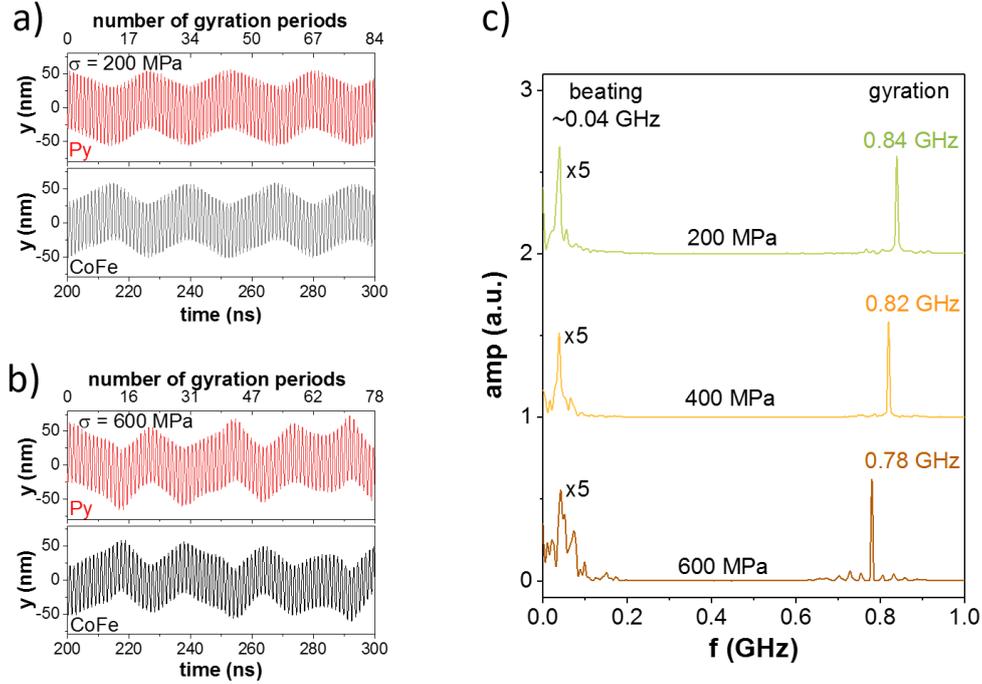

Fig. 6. (a,b) Time traces of the y-coordinate of the resonant vortex core trajectories of Fig. 5(b) taken for $\sigma_x$ = 200 MPa (a) and $\sigma_x$ = 600 MPa (b). (c) FFT spectra of the time traces of the y-coordinate as a function of stress.

Notably, the observed effects of the trajectories separation and beatings are present only for thin NM spacer ($t_{NM} \leq 20$ nm) and for the $c_1 c_2 = 1$ case, i.e. when the stacked vortices have the same in-plane circulations. This "stochastic repulsion" of the trajectories appears due to the increased interaction between the stress-induced magnetoelastic anisotropy and the strong magnetostatic coupling between two closely stacked magnetic vortices in the repulsion potential regime $(c_1 c_2; p_1 p_2) = (1; 1)$. Such complex dynamics of the stacked vortex pair modulated by the stress-induced magnetoelastic anisotropy allows controlling not only the gyration frequencies but also the lateral separation of the vortex core trajectories. This leads to the non-zero net magnetization change over the single gyration period, thus allowing the magnetoresistive readout of the in-phase gyration mode, which is often elusive when no analyzer layer is used.

### 4. Conclusions and outlook

In this paper, we put the focus on the dynamics of the MS/NM/nMS double-vortex structure with different $(c; p)$ combinations as a function of the uniaxial stress and showed that a complex dynamics of the coupled vortices occurs when the interlayer magnetostatic coupling is interfered by the magnetoelastic anisotropy. The micromagnetic simulation result reveal, that, depending on the vortex core polarities, the effects of the magnetoelastic anisotropy are qualitatively different and lead to the synchronization of the gyration frequencies for the case of parallel polarities, or to the lateral separation of the gyration trajectories for the case of anti-parallel polarities. In this paper, we considered the double-vortex structure with one magnetostrictive layer. However, when the double-vortex structure consists of two magnetostrictive layers with the magnetostriction constants of the opposite sign (e.g. CoFe ($\lambda_s > 0$) and Ni ($\lambda_s < 0$)), more complex stress-induced behavior is expected, which is beyond of the scope of this paper.

The experimental detection of the effects predicted here can be performed, for example, by measuring the microwave absorption spectra of the double-vortex structures excited by a microwave Oersted field from the adjacent stripline antenna. Another detection method consists of passing the rf current laterally through the structure using contact pads and measuring the microwave rectification voltage resulting from mixing of the oscillating magnetoresistance and injected rf current [26,27]. The latter method is considered more efficient due to the selective addressing and presence of an additional spin transfer torque acting on the vortex core. However, it may also introduce the unwanted Joule heating and non-linear effects due to large driving force hindering the efficient excitation of the vortex cores in such tri-layer structure. The initial states with different mutual orientations of the core polarities can be initiated on-demand by proper sequence of out-of-plane magnetic fields applied to the tri-layer and/or dc electrical currents passed vertically through the structure [5].

The stress-induced magnetoelastic anisotropy can be introduced via converse piezoelectric effect when the MS/NM/nMS tri-layer is grown and patterned on a ferroelectric relaxor substrate with large piezoelectric coefficients. For example, for commercially available lead magnesium niobate – lead titanate (Pb[Mg$_{1/3}$Nb$_{2/3}$)$_x$Ti$_{1-x}$]O$_3$, PMN-PT) single crystals the piezoelectric coefficient can be as high as 1500 ppm [28]. Considering the Young's modulus for CoFe Y = 250 GPa [29], the corresponding electric fields needed to obtain the stress values used in this work (up to 600 MPa) are close to 1 MV/m which can be easily achieved by proper electrodes designing.

**Acknowledgments**

The research leading to this result has been partly supported by the EU project TRANSPIRE (grant agreement No. 737038 from the EU Framework Programme for Research and Innovation HORIZON 2020.

**Supplementary data**

Fig. S1 shows the resonant frequencies (a) and resonant trajectories (b) of the simulated stress-modified dynamics in CoFe(10)NM(160)Py(20) double vortex structure for the $(c_1c_2; p_1p_2) = (-1; -1)$ case. Since the spacer is 160 nm thick, the vortices gyrate independently and no significant influence of the magnetostatic coupling on the dynamics is observed.

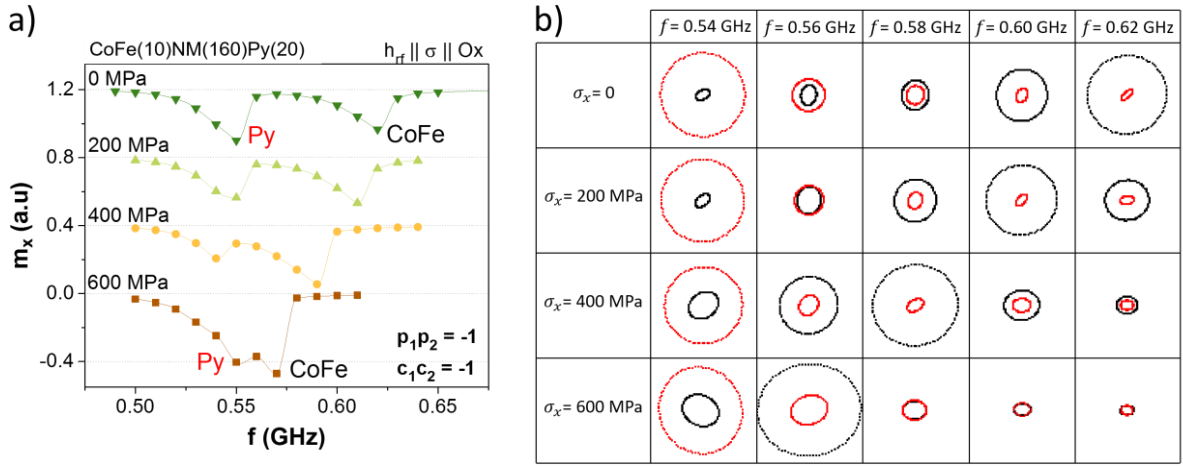

Fig. S1. (a) Simulated spectra of the 300 nm diameter CoFe(10)NM(160)Py(20) double-vortex structure for the $(c_1c_2; p_1p_2) = (-1; -1)$ case taken at different uniaxial tensile stress $\sigma_x$. The plots are vertically offset for better visibility. (b) Trajectories of the CoFe (black) and Py (red) vortex cores recorded for excitation frequencies $f$ = 0.54; 0.56; 0.58; 0.60 and 0.62 GHz without applied stress ($\sigma_x = 0$) and for $\sigma_x$ = 200; 400 and 600 MPa. The table cell size corresponds to the square area of 80×80 nm² centered in the middle of the disk.

Fig. S2 shows the resonant frequencies (a) and resonant trajectories (b) of the simulated stress-modified dynamics in CoFe(10)NM(60)Py(20) double vortex structure for the $(c_1 c_2; p_1 p_2) = (1; 1)$ case. For moderate magnetostatic coupling, the CoFe and Py vortex core trajectories are concentric.

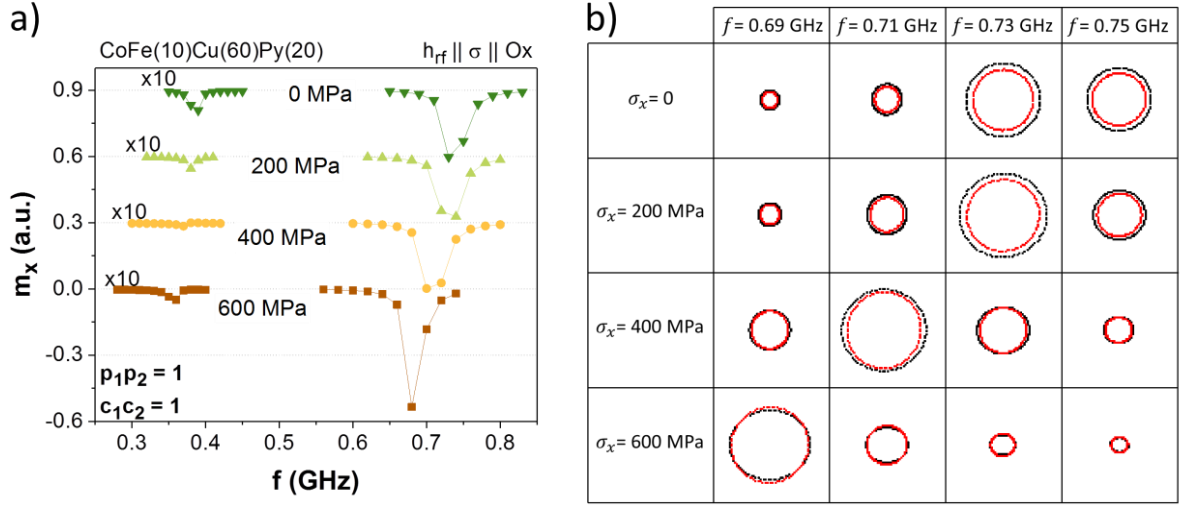

Fig. S2. (a) Simulated spectra of the 300 nm diameter CoFe(10)NM(60)Py(20) double-vortex structure for the $(c_1 c_2; p_1 p_2) = (1; 1)$ case taken at different uniaxial tensile stress $\sigma_x$. The plots are vertically offset for better visibility. The amplitudes of the low-frequency branch are multiplied by a factor of 10. (b) Trajectories of the CoFe (black) and Py (red) vortex cores recorded for excitation frequencies $f$ = 0.69; 0.71; 0.73 and 0.75 GHz without applied stress ($\sigma_x = 0$) and for $\sigma_x$ = 200; 400 and 600 MPa. The table cell size corresponds to the square area of 80×80 nm² centered in the middle of the disk.